\documentclass[%
 reprint,
 amsmath,amssymb,
 aps,
 prl,
 superscriptaddress,
 nobibnotes,
 floatfix,
]{revtex4-2}

\usepackage{graphicx}
\usepackage{dcolumn}
\usepackage{bm}
\usepackage{flushend}%

\makeatletter
\AtBeginDocument{%
  \def\bibsection{%
    \par
    \onecolumngrid@push
    \begingroup
      \baselineskip26\p@
      \bib@device{\textwidth}{245.5\p@}%
    \endgroup
    \nobreak\@nobreaktrue
    \addvspace{19\p@}%
    \par
    \onecolumngrid@pop
  }%
}
\makeatother

\usepackage[colorlinks=false, hidelinks]{hyperref}

\begin{document}

\title{Exploring Physical Mechanisms Underlying the Emergence of Collective Order}

\newcommand{\WmuAffiliation}{Postgraduate training base Alliance of Wenzhou Medical University, Wenzhou, Zhejiang 325000, China}
\newcommand{\WenzhouInstituteAffiliation}{Center for Theoretical Interdisciplinary Sciences, Wenzhou Institute, University of Chinese Academy of Sciences, Wenzhou, Zhejiang 325001, China}
\newcommand{\SoftMatterLabAffiliation}{Zhejiang Key Laboratory of Soft Matter Biomedical Materials, Wenzhou Institute, University of Chinese Academy of Sciences, Wenzhou, Zhejiang 325000, China}

\author{Yitong Cai}
\affiliation{\WmuAffiliation}
\affiliation{\WenzhouInstituteAffiliation}
\affiliation{\SoftMatterLabAffiliation}

\author{Ruizhe Yan}
\affiliation{\WenzhouInstituteAffiliation}
\affiliation{\SoftMatterLabAffiliation}
\affiliation{School of Medicine and Health, Harbin Institute of Technology, Harbin, 150001 China}

\author{Dongrun Jian}
\affiliation{\WenzhouInstituteAffiliation}
\affiliation{\SoftMatterLabAffiliation}
\affiliation{Department of Physics, Nanjing University, Nanjing, Jiangsu 210093, China}

\author{Jie Su}
\email{sj0410@ucas.ac.cn}
\affiliation{\WenzhouInstituteAffiliation}
\affiliation{\SoftMatterLabAffiliation}

\author{Jin Wang}
\email{jin.wang.1@stonybrook.edu}
\affiliation{\WmuAffiliation}
\affiliation{\WenzhouInstituteAffiliation}
\affiliation{\SoftMatterLabAffiliation}
\affiliation{Department of Chemistry and of Physics and Astronomy, State University of New York at Stony Brook, Stony Brook, New York 11794, USA}

\begin{abstract}
Emergence of collective order from disordered constituents is a hallmark of active matter, and the Vicsek model (VM) serves as a paradigmatic system for its study. However, whether density-induced emergence in the VM differs between angular (intrinsic) and vectorial (extrinsic) noise, and what mechanisms govern the transitions, remain largely unexplored. Remarkably, we uncover that collective order emerges from a disordered state in a continuous-like manner under intrinsic noise but abruptly under extrinsic noise. Combining nonequilibrium landscape-flux theory with a coarse-grained mapping method, we show that the landscape evolves continuously from a single basin to two and back to one under intrinsic noise, but the two-basin landscape appears abruptly under extrinsic noise across the critical transition to flocking. The counterclockwise nonequilibrium flux sustains global circular motion, providing the driving force for reshaping the landscape. The mean flux and entropy production rate change steeply near the emergence threshold and exhibit pronounced peaks within the coexistence region, identifying them as the dynamical and thermodynamic origins of emergence. Our findings reveal a distinct abrupt emergence under extrinsic noise and offer deeper mechanistic insight, practical applications (e.g., locating phase boundaries and distinguishing transition types), and extensibility to other nonequilibrium collective systems.
\end{abstract}

\maketitle

Active matter systems are ubiquitous across scales in nature, ranging from microscopic bacteria and artificial Janus colloids to macroscopic fish schools, bird flocks, and mammalian herds~\cite{Ramaswamy2010, Bechinger2016, Marchetti2013}. Comprising self-propelled units that consume energy to generate mechanical motion, these systems break detailed balance at the microscopic scale and exhibit abundant and novel collective behaviors, such as dynamic chirality, vortex lattices, and motility-induced phase separation~\cite{Sumino2012, Bricard2015, Buttinoni2013, Cates2015, Tailleur2008}. Of particular interest is \emph{emergence}, the spontaneous appearance of macroscopic order from microscopically disordered constituents~\cite{Anderson1972}, a key to understanding how complex collective behavior arises from simple individual interactions in nonequilibrium systems. Specifically, individual active particles always exhibit random or disordered motion, yet upon crossing a threshold density, the system spontaneously orders itself into a collective state~\cite{Halley2008}. Understanding this density-induced emergence provides a pathway to unraveling how crowding governs collective order across complex nonequilibrium systems~\cite{Buhl2006, Bricard2013, Anderson2024}, yet elucidating its underlying mechanisms remains a central challenge.

To better understand the emergent behavior, we introduce the Vicsek model (VM), a simplified model capturing the essence of collective motion, which can reproduce the ordered swarming observed in natural systems such as bird flocks and insect swarms~\cite{Vicsek1995}. The original VM study, which incorporates alignment interactions and angular noise (intrinsic noise) acting on a particle's own response, shows that the system exhibits ordered collective motion at low noise intensity and high density but remains disordered at high noise intensity or low density. The study also reveals a sigmoid-shaped density-induced transition from disorder to order, establishing the VM as a paradigmatic system for studying emergence. Since then, the VM has been extended into numerous variants that exhibit diverse collective motions~\cite{Shaebani2020, Farrell2012}, such as vortex lattices~\cite{Nagai2015} and rotating droplets~\cite{Liebchen2017}. Among these, one particularly interesting variant employs vectorial noise (extrinsic noise) acting on the signal a particle receives from its environment~\cite{Gregoire2004, Pimentel2008, Baglietto2014, Gulich2018}. Previous studies of the noise-induced loss of collective order report that intrinsic noise yields a continuous-like transition, whereas extrinsic noise produces a sharper, discontinuous one~\cite{Gregoire2004, Pimentel2008, Chate2008, Baglietto2009, Baglietto2012, Ginelli2016, Cavagna2021, Chate2020, Yan2025}. These studies are all confined to noise-induced transitions at a fixed density, while the density-driven ordering, i.e., the emergent behavior, has received far less attention. It therefore remains an open and important question whether the emergent behavior in VM systems differs between intrinsic and extrinsic noise, and what the underlying transition mechanisms are.

In this paper, we uncover an abrupt emergence in the VM under extrinsic noise, rather than a continuous-like one under intrinsic noise. By combining landscape-flux theory~\cite{Wang2008,Wang2015,Fang2019,LiWang2014,Yan2013,Xu2021}, widely applied in biophysics, ecology, and active matter, with a coarse-grained mapping method~\cite{Su2024,Yan2025,Jian2025}, we project the VM system onto the local density and order degree phase space, and subsequently quantify the potential landscape and nonequilibrium flux in this phase space. During the density-induced transition, the landscape evolves continuously from a single basin (disorder phase) to two coexisting basins (band phase, consisting of a dense ordered band and a dilute disordered gas) and back to a single basin (order phase) under intrinsic noise. For extrinsic noise, the landscape shape changes discontinuously, with the intermediate-state potential basins appearing abruptly, signaling a different transition type. In contrast to the stabilizing landscape gradient, a counterclockwise nonequilibrium flux sustains global circular motion and acts as the driving force for reshaping the landscape. We further quantify the mean flux and entropy production rate, and find that both vary steeply near the emergence threshold and develop pronounced peaks within the coexistence region, revealing the dynamical and thermodynamic origins of emergence, respectively. Meanwhile, these changes and peaks are more pronounced under extrinsic noise, showing the discontinuous nature of the transition. Moreover, we calculate the cross-correlation asymmetry and the optimal transition paths, finding stronger time-reversal symmetry breaking within the band phase region, especially under extrinsic noise. We finally construct the $(\rho_0,\eta)$ phase diagrams for both noise types and verify the robustness of our approach across noise intensities.

\paragraph*{Model and methods.--}

The VM describes $N$ particles moving at constant speed $v_0$ in a two-dimensional system of size $L \times L$ with periodic boundary conditions~\cite{Vicsek1995}. Particle $i$ at position $\mathbf{r}_i(t)$ and orientation $\theta_i(t)$ obeys the discrete-time dynamics with time step $\Delta t$
\begin{align}
\mathbf{r}_i(t+\Delta t) &= \mathbf{r}_i(t) + \mathbf{v}_i(t)\Delta t, \label{eq:pos}\\
\theta_i(t+\Delta t) &= \operatorname{Arg}\bigl[\sum_{j \in \partial_i} \mathbf{v}_j(t)\bigr] + \eta_{\mathrm{in}}\xi_i, \label{eq:in}\\
\theta_i(t+\Delta t) &= \operatorname{Arg}\bigl[\sum_{j \in \partial_i} \left(\mathbf{v}_j(t) + \eta_{\mathrm{ex}}\boldsymbol{\xi}_i\right)\bigr], \label{eq:ex}
\end{align}
where $\mathbf{v}_i(t) = v_0(\cos\theta_i(t), \sin\theta_i(t))$ is the velocity of particle $i$. $\operatorname{Arg}[\cdot]$ denotes the orientation of the vector sum inside the brackets. At each time step, the orientation $\theta_i$ is updated by alignment with neighbors $j \in \partial_i$ within radius $R$ under intrinsic noise (Eq.~\eqref{eq:in}) or extrinsic noise (Eq.~\eqref{eq:ex}). The former is angular noise, with $\xi_i$ uniformly distributed in $[-\pi, \pi]$ and intensity $\eta_{\mathrm{in}}$, while the latter is vectorial noise, with intensity $\eta_{\mathrm{ex}}$ acting on a random unit vector $\boldsymbol{\xi}_i$.

\begin{figure}[t]
\centering
\includegraphics[width=\columnwidth]{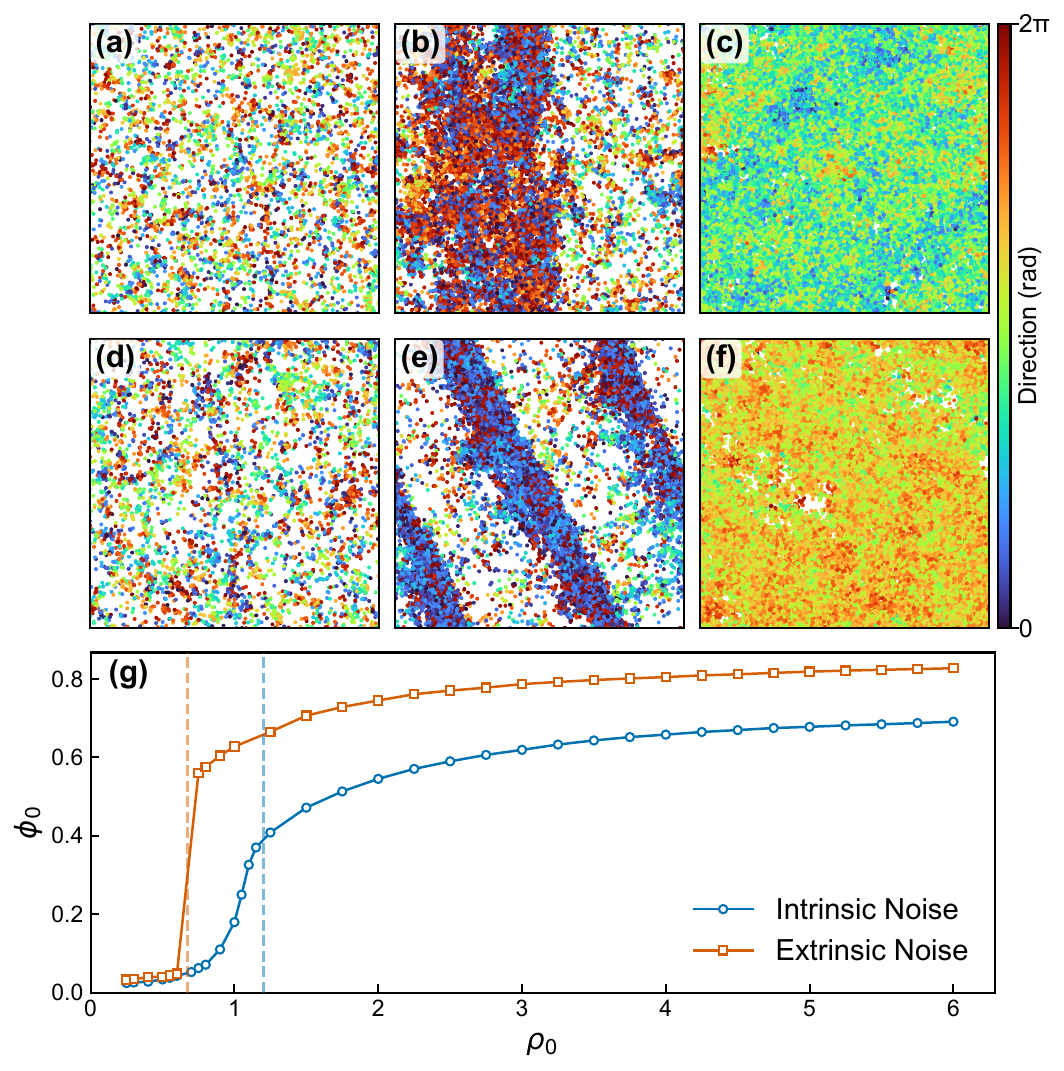}
\caption{Density-induced disorder-order transition in the VM under different types of noise. (a--f) Typical steady-state snapshots under intrinsic (a--c) and extrinsic noise (d--f). The total densities are: (a,d) $\rho_0=0.5$ (disordered); (b) $\rho_0=1.75$, (e) $\rho_0=2.0$ (band); (c,f) $\rho_0=5.0$ (ordered). (g) Total order degree $\phi_0$ as a function of total density $\rho_0$ for intrinsic noise and extrinsic noise. The dashed lines mark the onset of the band phase region under intrinsic and extrinsic noise.}
\label{fig:snapshots}
\end{figure}

In our simulations, we take the interaction radius $R$ and the time step $\Delta t$ as the basic units for length and time, respectively, i.e., $R = 1$ and $\Delta t = 1$. We set $\eta_{\mathrm{in}} = 0.4$, $\eta_{\mathrm{ex}} = 0.6$, $v_0 = 0.5$, and $L = 100$, while varying the total density $\rho_0 = N/L^2$ by changing $N$. The first $10^5$ time steps are discarded for relaxation, and data are sampled over the subsequent $10^5$ steps, with ensemble averages taken over 10 independent runs.

To uncover the underlying transition mechanisms of VM systems, we first apply the coarse-grained mapping method~\cite{Su2024,Yan2025,Jian2025,Bolley2012} that projects the particle dynamics onto a phase space spanned by the local density $\rho = N_m/A_m$ and local order degree $\phi = |\sum_{i \in m}\mathbf{v}_i|/(N_m v_0)$ of each cell with area $A_m$ containing $N_m$ particles (see Supplemental Material (SM) for details), and then obtain the corresponding stochastic Langevin equations:
\begin{equation}
\begin{aligned}
\dot\rho &= F_\rho(\rho,\phi) + \xi_\rho(t),\\
\dot\phi &= F_\phi(\rho,\phi) + \xi_\phi(t).
\end{aligned}
\end{equation}
Here, $\mathbf{F}=(F_\rho,F_\phi)^{\mathrm{T}}$ is the deterministic driving force and $\boldsymbol{\xi}=(\xi_\rho,\xi_\phi)^{\mathrm{T}}$ is the stochastic force satisfying $\left\langle\boldsymbol{\xi}(t)\boldsymbol{\xi}^{\mathrm{T}}(t')\right\rangle=2\mathbf{D}(\rho,\phi)\delta(t-t')$, where $\mathbf{D}$ is the diffusion tensor (see details in SM).

Subsequently, the landscape-flux framework~\cite{Wang2008,Wang2015,Fang2019,WuWang2013,Wu2021} (see SM for details) yields the nonequilibrium potential landscape and steady-state probability flux in this local $(\rho,\phi)$ phase space as
\begin{equation}
\begin{aligned}
U(\rho,\phi) &= -\ln P_{\mathrm{ss}}(\rho,\phi),\\
\mathbf{J}_{\mathrm{ss}}(\rho,\phi) &= \mathbf{F}(\rho,\phi)P_{\mathrm{ss}}(\rho,\phi)
- \nabla\cdot\left[\mathbf{D}(\rho,\phi)P_{\mathrm{ss}}(\rho,\phi)\right].
\end{aligned}
\label{eq:landscape_flux}
\end{equation}
The landscape and flux fields compactly characterize state stability and nonequilibrium circulation, respectively, thereby underpinning the dynamical and thermodynamic analyses below.

\paragraph*{Emergence under intrinsic and extrinsic noise.--}

Particle-configuration snapshots of the VM under intrinsic and extrinsic noise show the density-induced transition from disorder to collective order as the total density $\rho_0$ increases (Fig.~\ref{fig:snapshots}). At low density ($\rho_0=0.5$), both intrinsic and extrinsic noise lead to globally disordered configurations (Fig.~\ref{fig:snapshots}a,d). At intermediate density, a band phase region appears, in which a dense ordered band propagates through and coexists with a dilute disordered gas~\cite{Kursten2020, Mangeat2020, Clusella2021}. The band is narrower, more compact, and more sharply defined under extrinsic noise than under intrinsic noise (Fig.~\ref{fig:snapshots}e,b), indicating stronger separation between the two phases and a more abrupt density-induced disorder-order transition. At high density ($\rho_0=5.0$), both systems enter a highly ordered state, with most particles moving collectively in a common direction (Fig.~\ref{fig:snapshots}c,f).

The total order degree $\phi_0 = |\sum_i \mathbf{v}_i|/(Nv_0)$ then quantifies the density-induced disorder-order transition (Fig.~\ref{fig:snapshots}g). Under intrinsic noise, $\phi_0$ increases continuously with density, showing a smooth, finite-slope rise around $\rho_0 \approx 1.25$. However, under extrinsic noise, $\phi_0$ increases abruptly, with a sharp, step-like change around $\rho_0 \approx 0.75$. Together, the particle-configuration snapshots and $\phi_0$-$\rho_0$ curves establish the macroscopic phenomenology of emergence, revealing an intermediate band phase region and continuous-like or abrupt transitions under the two noise types.

\paragraph*{Potential landscape and flux of emergence.--}

\begin{figure}[t]
\centering
\includegraphics[width=\columnwidth]{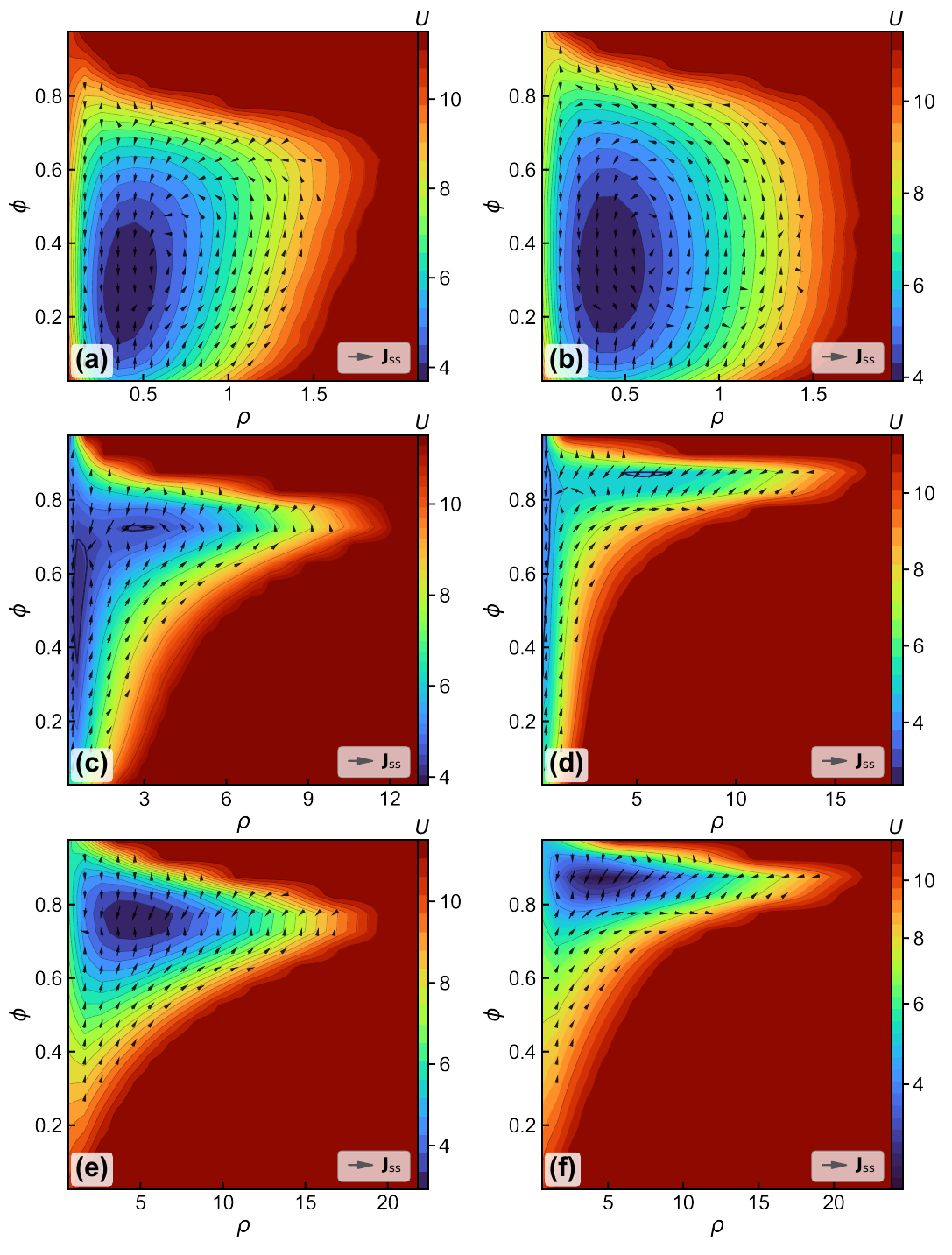}
\caption{Nonequilibrium potential landscape $U$ (color) and steady-state probability flux $\mathbf{J}_{\mathrm{ss}}$ (black arrows) in the local $(\rho,\phi)$ phase space under intrinsic noise (left column) and extrinsic noise (right column). The total densities are: (a,b) $\rho_0=0.5$ (disordered); (c) $\rho_0=1.75$, (d) $\rho_0=2.0$ (band); (e,f) $\rho_0=5.0$ (ordered).}
\label{fig:U_J}
\end{figure}

To further uncover the mechanism underlying the emergent behavior in the VM systems, we use the coarse-grained mapping method to project the dynamical trajectories of Vicsek particles in position space onto the local $(\rho,\phi)$ phase space. Then, by applying the landscape-flux theory in this phase space, we quantify the nonequilibrium potential landscape $U$ and steady-state probability flux $\mathbf{J}_{\mathrm{ss}}$ for both noise types at various $\rho_0$ (Fig.~\ref{fig:U_J}).

Under both intrinsic and extrinsic noise, increasing $\rho_0$ takes the landscape from a single disordered basin located in the low-$\phi$ region to two coexisting basins and finally to a single ordered basin located in the high-$\phi$ region. The coexisting basins located in the low-$\phi$ and high-$\phi$ regions, respectively, correspond to the dilute disordered gas and the dense ordered band (Fig.~\ref{fig:U_J}c,d). The density-resolved landscape sequence reveals two distinct transition types (see Fig. S3 in SM). Under intrinsic noise, the landscape evolves continuously from a single basin to two and back to one, representing a continuous-like transition. By contrast, under extrinsic noise, the two-basin landscape appears abruptly, implying an abrupt transition. These landscape evolution results are consistent with the $\phi_0$-$\rho_0$ curves in Fig.~\ref{fig:snapshots}g.

Moreover, we observe that the nonzero flux circulates counterclockwise in the local phase space, revealing detailed-balance breaking beyond the landscape description (black arrows in Fig.~\ref{fig:U_J}). Different from the landscape gradient, which stabilizes the local states, the rotational flux drives the system away from simple localization and sustains global circular movements, thereby providing a driving force for the change of the landscape shape. Within the coexistence region, the band moves persistently in an ordered and directed manner, so that each position-space cell alternately samples the dense ordered band and the dilute disordered gas, a process that corresponds to the switching between two potential basins in the phase space. Unlike in equilibrium, where the switching between basins relies solely on noise, under nonequilibrium conditions the nonequilibrium flux can dominate and drive such switching. In short, the interplay of the potential landscape and the nonequilibrium flux provides the dynamical mechanism for the density-induced emergent behavior in the VM systems.

\begin{figure}[t]
\centering
\includegraphics[width=\columnwidth]{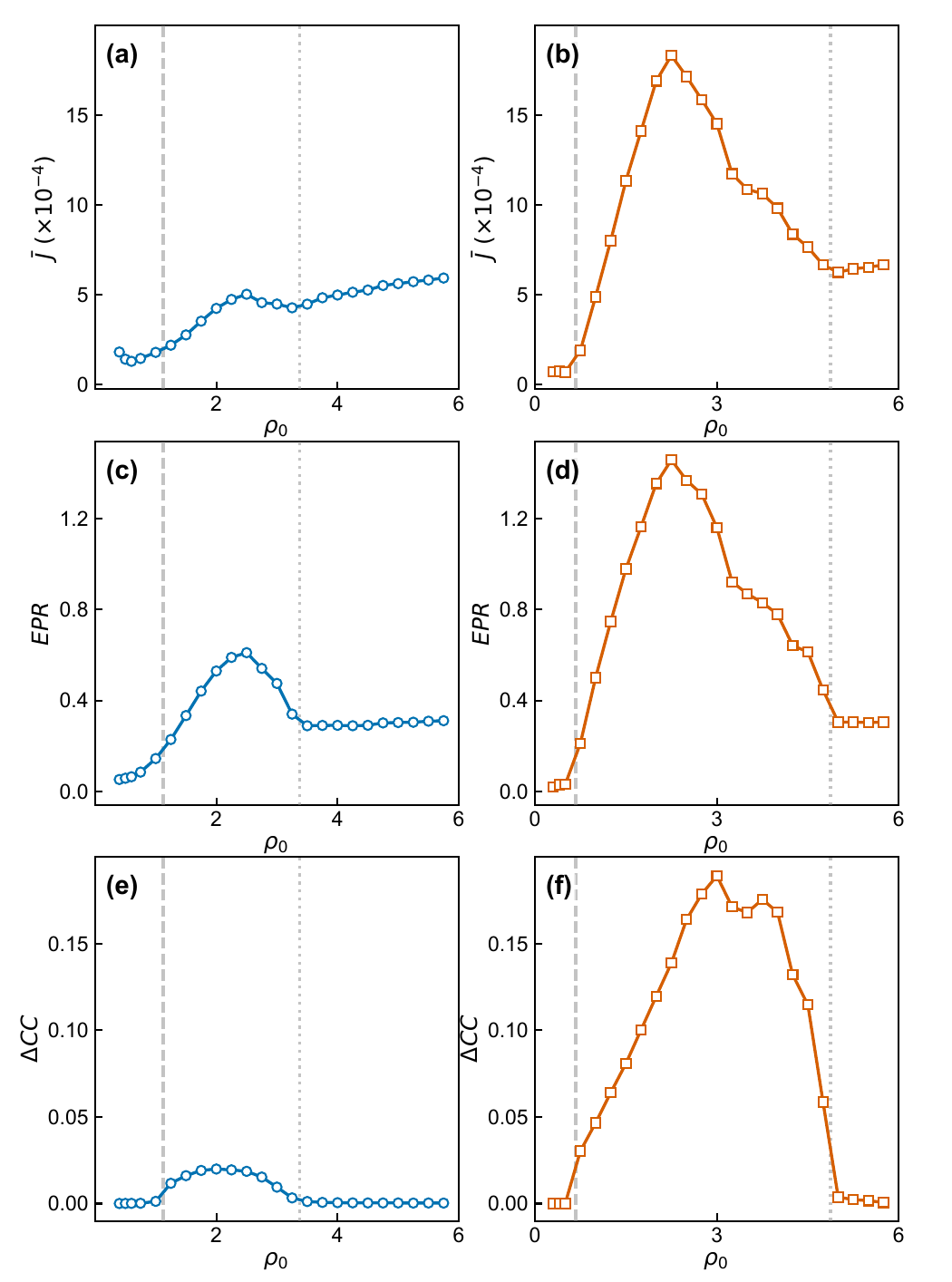}
\caption{Dynamical, thermodynamic, and time-reversal signatures of emergence under intrinsic noise (left column) and extrinsic noise (right column). (a,b) The mean flux magnitude $\bar J$, (c,d) entropy production rate $EPR$, and (e,f) cross-correlation asymmetry $\Delta CC$ as functions of the total density $\rho_0$. Gray dashed and dotted lines mark the left and right boundaries of the band phase region, respectively.}
\label{fig:indicators}
\end{figure}

\begin{figure}[t]
\centering
\includegraphics[width=\columnwidth]{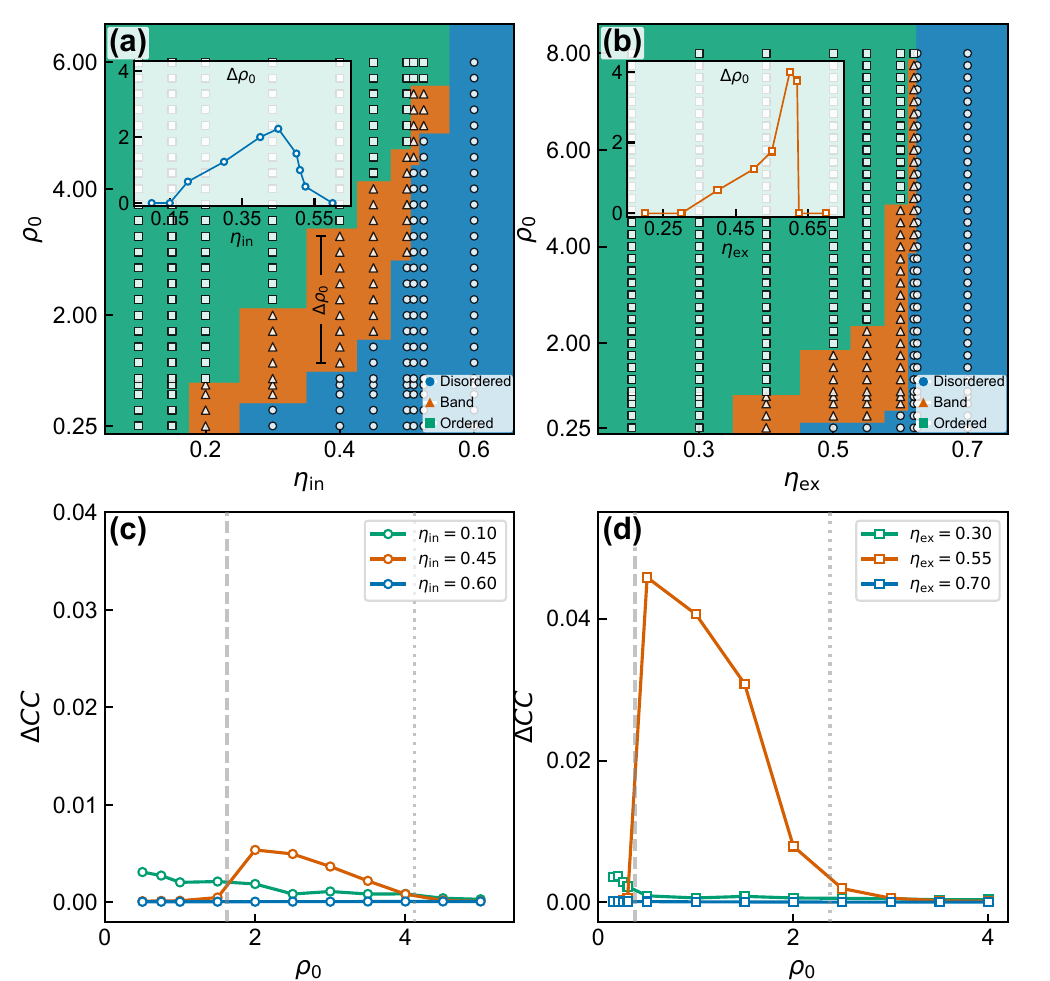}
\caption{Phase diagram and robustness of this approach across noise intensities. (a,b) Phase diagrams in the $(\rho_0,\eta)$ plane, consisting of disordered (blue), band (orange), and ordered (green) phases. Insets show the density width $\Delta\rho_0$ of the band phase region as a function of $\eta$. (c,d) $\Delta CC$ profiles as functions of $\rho_0$ at representative $\eta$ for (c) intrinsic and (d) extrinsic noise.}
\label{fig:phase_diagram}
\end{figure}

\paragraph*{Dynamical and thermodynamic origins of emergent behavior.--}

We next quantify the dynamical, thermodynamic, and irreversible aspects of emergence, i.e., the density-induced disorder-order transition in VM systems, by calculating the mean flux magnitude $\bar J$, entropy production rate $EPR$, and cross-correlation asymmetry $\Delta CC$ based on the landscape-flux theory~\cite{Seifert2012, Esposito2012} (see expressions and details in SM).

We observe that $\bar J$ and $EPR$ follow a similar trend under both intrinsic and extrinsic noise (Fig.~\ref{fig:indicators}a--d). While the system remains disordered, both observables rise slightly with $\rho_0$. Once the system crosses the left boundary of the band phase region, i.e., the emergence threshold of collective order, both observables rise steeply and exhibit a peak inside the band phase region (dashed lines in Fig.~\ref{fig:indicators}, consistent with the dashed lines in Fig.~\ref{fig:snapshots}g). After the peak, both observables decrease overall as $\rho_0$ increases until the right boundary (dotted line) is reached, beyond which they rise slightly again. Both phase boundaries thus exhibit clear slope changes, with a steepening at the left boundary and a leveling off at the right boundary. Consequently, the pronounced peak is confined to the coexistence region, while the single-phase regions remain monotonic. The steep variations near the transition threshold and the appearance of peaks in the band phase demonstrate that the nonequilibrium flux and $EPR$ are the dynamical and thermodynamic origins of the density-induced emergent behavior in VM systems, respectively~\cite{XuWang2020, FangWang2020}. Moreover, we find that under extrinsic noise, the change near the emergence threshold is steeper and the peak in the band phase is markedly higher, consistent with the two transition types in Fig.~\ref{fig:snapshots}g.

In addition, we find that $\Delta CC$ first increases and then decreases within the band phase region, developing a peak, which reflects stronger time-reversal asymmetry in the band phase~\cite{Borthne2020} (Fig.~\ref{fig:indicators}e,f). Compared with the intrinsic-noise case, the peak under extrinsic noise is markedly larger, also indicating stronger time-reversal asymmetry. Furthermore, the non-overlapping forward and backward optimal transition paths provide a trajectory-level signature of this time-reversal asymmetry in the band phase region~\cite{Wang2010}. Their non-overlap is more pronounced under extrinsic noise (see SM for details). In short, the changes in $\bar J$, $EPR$, and $\Delta CC$ not only reflect the physical nature of the system, but also provide practical applications such as predicting or giving early warnings of transition boundaries or bifurcations, and estimating transition types.

\paragraph*{Phase diagram and robustness of this approach.--}

Finally, we perform additional and extensive numerical simulations to explore the emergent behaviors of the VM system under various noise intensities $\eta$, thereby obtaining phase diagrams in the $(\rho_0,\eta)$ plane (Fig.~\ref{fig:phase_diagram}a,b). Under both intrinsic and extrinsic noise, the phase diagrams share a common structure. At small $\eta$, the system is ordered (green) throughout; at intermediate $\eta$, it proceeds from disordered (blue) to band (orange) to ordered (green) with increasing density; and at sufficiently large $\eta$, it remains disordered (blue). Despite this common structure, the disorder-order transition remains continuous-like under intrinsic noise but is discontinuous under extrinsic noise (see Fig. S6 in SM). Beyond the transition type, the band phase exhibits different shapes in the phase diagram under intrinsic and extrinsic noise. To characterize how the band phase region varies across the parameter space, we calculate its density width $\Delta\rho_0$ and show its dependence on $\eta$ in the insets of Fig.~\ref{fig:phase_diagram}a,b. In contrast to intrinsic noise, where $\Delta\rho_0$ increases first and then decreases smoothly as $\eta$ increases, under extrinsic noise the $\Delta\rho_0$ peak is higher and sharper, indicating that the band phase region vanishes abruptly within a narrow $\eta$ interval.

To verify the robustness of this approach across noise intensities, we calculate the $\Delta CC$-$\rho_0$ curves at three representative $\eta$ as examples. As shown in Fig.~\ref{fig:phase_diagram}(c,d), we observe that under both intrinsic and extrinsic noise, $\Delta CC$ exhibits a pronounced peak at intermediate $\eta$, with the peak located inside the band phase region. By contrast, no such peak appears at low or high $\eta$, where the system remains ordered or disordered throughout, respectively. The correspondence between these distinct $\Delta CC$ profiles and the configuration-based phase diagrams shows that $\Delta CC$ can be used to identify the various phases and to detect whether a phase transition has occurred, demonstrating the robustness of this approach. Similar results are observed for other physical quantities, such as the $EPR$-$\rho_0$ curves in Fig. S7 in SM.

\paragraph*{Conclusion.--}

In summary, compared with previous findings that collective order emerges in a continuous-like manner under intrinsic noise in the VM, we find that extrinsic noise induces an abrupt emergence. By combining landscape-flux theory with the coarse-grained mapping method, we explore the underlying mechanisms of the emergent behaviors under both noise types. The nonequilibrium landscape evolves continuously from a single basin to two and back to one under intrinsic noise, while the two-basin landscape appears abruptly under extrinsic noise, consistent with the corresponding macroscopic transition types. Different from the landscape gradient stabilizing the local states, the counterclockwise nonequilibrium flux drives the system toward global circular motion and provides a driving force for reshaping the landscape. We further quantify the mean flux and $EPR$, both of which change steeply near the emergence threshold and show pronounced peaks within the band phase, demonstrating that the flux and $EPR$ are respectively the dynamical and thermodynamic origins of the critical transition or bifurcation. More interestingly, these features are more abrupt and pronounced under extrinsic noise, representing a more abrupt disorder-order phase transition. Additionally, we construct the $(\rho_0,\eta)$ phase diagrams for both noise types and verify the robustness of our approach across noise intensities. Our findings show a distinct abrupt emergence of collective order under extrinsic noise in VM systems, not only providing deeper insight into the physical mechanisms of emergent behaviors, but also offering practical applications, such as locating phase boundaries or bifurcations, and distinguishing transition types.

Y. Cai, R. Yan, D. Jian, and J. Su gratefully acknowledge funding from National Natural Science Foundation of China (Nos. 12404237 and 12234019), the Zhejiang Key Laboratory of Soft Matter Biomedical Materials (Nos. 2025ZY01036 and 2025E10072) and Wenzhou Institute of Biomaterials and Engineering (No. WIUCASQD2022012).

\bibliography{ref.bib}

\end{document}


\title{Supplemental Material: Exploring Physical Mechanisms Underlying the Emergence of Collective Order}

\newcommand{\WmuAffiliation}{Postgraduate training base Alliance of Wenzhou Medical University, Wenzhou, Zhejiang 325000, China}
\newcommand{\WenzhouInstituteAffiliation}{Center for Theoretical Interdisciplinary Sciences, Wenzhou Institute, University of Chinese Academy of Sciences, Wenzhou, Zhejiang 325001, China}
\newcommand{\SoftMatterLabAffiliation}{Zhejiang Key Laboratory of Soft Matter Biomedical Materials, Wenzhou Institute, University of Chinese Academy of Sciences, Wenzhou, Zhejiang 325000, China}

\author{Yitong Cai}
\affiliation{\WmuAffiliation}
\affiliation{\WenzhouInstituteAffiliation}
\affiliation{\SoftMatterLabAffiliation}

\author{Ruizhe Yan}
\affiliation{\WenzhouInstituteAffiliation}
\affiliation{\SoftMatterLabAffiliation}
\affiliation{School of Medicine and Health, Harbin Institute of Technology, Harbin, 150001 China}

\author{Dongrun Jian}
\affiliation{\WenzhouInstituteAffiliation}
\affiliation{\SoftMatterLabAffiliation}
\affiliation{Department of Physics, Nanjing University, Nanjing, Jiangsu 210093, China}

\author{Jie Su}
\email{sj0410@ucas.ac.cn}
\affiliation{\WenzhouInstituteAffiliation}
\affiliation{\SoftMatterLabAffiliation}

\author{Jin Wang}
\email{jin.wang.1@stonybrook.edu}
\affiliation{\WmuAffiliation}
\affiliation{\WenzhouInstituteAffiliation}
\affiliation{\SoftMatterLabAffiliation}
\affiliation{Department of Chemistry and of Physics and Astronomy, State University of New York at Stony Brook, Stony Brook, New York 11794, USA}
\maketitle

\section{I.\ Model and simulation details}
\label{sec:simdetails}

The Vicsek model (VM) describes $N$ self-propelled particles moving at constant speed $v_0$ in a two-dimensional square box of size $L \times L$ with periodic boundary conditions~\cite{Vicsek1995}. Particle $i$ is specified by its position $\mathbf{r}_i(t)$ and orientation $\theta_i(t)$, and its velocity reads $\mathbf{v}_i(t) = v_0(\cos\theta_i(t), \sin\theta_i(t))$. The position updates as
\begin{equation}
\mathbf{r}_i(t+\Delta t) = \mathbf{r}_i(t) + \mathbf{v}_i(t)\,\Delta t,
\label{eq:SI_position}
\end{equation}
where $\Delta t$ is the discrete time step. At each time step the orientation is updated by aligning with the neighbors $j \in \partial_i$ within the interaction radius $R$, where $\partial_i = \{j : |\mathbf{r}_j - \mathbf{r}_i| \le R\}$ includes particle $i$ itself. The two noise types differ only in where the uncertainty enters the alignment rule~\cite{Pimentel2008, Gregoire2004}.

\paragraph*{Intrinsic noise (angular noise).}
Noise of intensity $\eta_{\mathrm{in}}$ perturbs the alignment direction after the averaging step and thus acts as an additive ``response error'':
\begin{equation}
\theta_i(t+\Delta t) = \operatorname{Arg}\left[\sum_{j \in \partial_i} \mathbf{v}_j(t)\right] + \eta_{\mathrm{in}}\,\xi_i,
\label{eq:SI_intrinsic}
\end{equation}
where $\xi_i$ is drawn independently for each particle and each time step from a uniform distribution on $[-\pi, \pi]$. Because the noise is added after the neighbor averaging, its effective strength is independent of the local order.

\paragraph*{Extrinsic noise (vectorial noise).}
Noise of intensity $\eta_{\mathrm{ex}}$ perturbs the velocity perceived by each particle before the averaging step and thus acts as a ``perception error'':
\begin{equation}
\theta_i(t+\Delta t) = \operatorname{Arg}\left[\sum_{j \in \partial_i} \left(\mathbf{v}_j(t) + \eta_{\mathrm{ex}}\,\boldsymbol{\xi}_i\right)\right],
\label{eq:SI_extrinsic}
\end{equation}
where $\boldsymbol{\xi}_i$ is a random unit vector with a uniformly distributed direction. Here the noise competes directly with the local alignment and therefore couples nonlinearly to the local order degree.

We use $R = 1$, $v_0 = 0.5$, $\Delta t = 1$, and $L = 100$. The total density $\rho_0 = N/L^2$ is tuned by changing the particle number $N$. The noise intensity is fixed at $\eta_{\mathrm{in}} = 0.4$ and $\eta_{\mathrm{ex}} = 0.6$ for the intrinsic- and extrinsic-noise analyses in the main text, and is varied to construct the $(\rho_0, \eta)$ phase diagrams. Starting from initial configurations with randomly distributed positions and orientations, the system is first run for $10^5$ time steps to relax into the steady state; configurations are then sampled over the subsequent $10^5$ steps. Ensemble averages are taken over 10 independent runs with different initial conditions.

\section{II.\ Coarse-grained mapping method}
\label{sec:cgm}

\begin{figure*}[t]
\centering
\includegraphics[width=\textwidth]{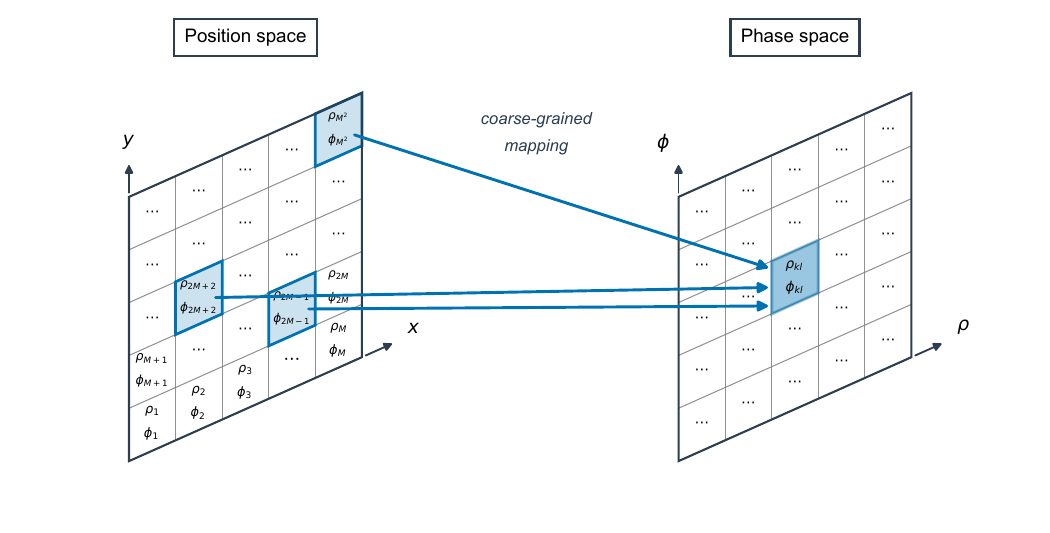}
\caption{Schematic of the coarse-grained mapping from the evolution of cells in position space to the local $(\rho,\phi)$ phase space. The cell area is adjusted with the total density to maintain a prescribed mean particle count per cell. A cell is mapped onto bin $(k,l)$ if $\rho_m(t) \in [k\delta\rho, (k+1)\delta\rho)$ and $\phi_m(t) \in [l\delta\phi, (l+1)\delta\phi)$.}
\label{fig:SI_mapping}
\end{figure*}

We coarse grain the microscopic particle dynamics onto a phase space spanned by the local density $\rho$ and the local order degree $\phi$, following the coarse-grained mapping method~\cite{Su2024, Yan2025, Jian2025}. These two order parameters are chosen because they quantify particle accumulation and directional alignment, respectively, which are essential for characterizing the emergence of collective order. To construct the mapping, the position space is covered by cells (Fig.~\ref{fig:SI_mapping}). Cell $m$ has area $A_m$ and contains $N_m$ particles. Each cell carries a local density $\rho_m = N_m/A_m$ and a local order degree $\phi_m = |\sum_{i \in m} \mathbf{v}_i|/(N_m v_0)$. Replacing the cell $m$ by the whole system in $\phi_m$ gives the total order degree $\phi_0 = |\sum_{i} \mathbf{v}_i|/(N v_0)$, which quantifies the density-induced disorder-order transition in the main text.

Each cell contributes one point $(\rho_m,\phi_m)$ at time $t$. We discretize the local $(\rho,\phi)$ phase space into bins of size $\delta\rho \times \delta\phi$ (Fig.~\ref{fig:SI_mapping}). At time $t$, cell $m$ is mapped onto bin $(k,l)$ if $\rho_m(t) \in [k\delta\rho, (k+1)\delta\rho)$ and $\phi_m(t) \in [l\delta\phi, (l+1)\delta\phi)$. The occupation probability $P(k,l,t)$ is the fraction of cells assigned to bin $(k,l)$ at time $t$. Binning all cells at each sampled time yields the time-dependent distribution $P(\rho,\phi,t)$. Its evolution is described by the Fokker--Planck equation as follows:
\begin{equation}
\frac{\partial P}{\partial t}
=
-\nabla\cdot\left[\mathbf F P-\nabla\cdot(\mathbf D P)\right].
\label{eq:FP}
\end{equation}

The deterministic driving force $\mathbf F$ in the Fokker--Planck equation is obtained from the trajectories of individual cells. Averaging the rates of change within each phase-space bin gives
\begin{equation}
F_\rho(\rho,\phi)=\left\langle\dot\rho_i(\rho,\phi,t)\right\rangle_{i,t},
\qquad
F_\phi(\rho,\phi)=\left\langle\dot\phi_i(\rho,\phi,t)\right\rangle_{i,t}.
\label{eq:F}
\end{equation}
The dots denote time derivatives. Within each simulation run, the averages are taken over cells $i$ and sampled times $t$ for which $(\rho_i(t),\phi_i(t))$ lies in the bin $(\rho,\phi)$.

In addition, the stochastic force
\[
\boldsymbol\xi_i(\rho,\phi,t)
=
\begin{pmatrix}
\dot\rho_i(\rho,\phi,t)-F_\rho(\rho,\phi)\\
\dot\phi_i(\rho,\phi,t)-F_\phi(\rho,\phi)
\end{pmatrix}
\]
represents the fluctuation of the local rates of change about the drift for cell $i$. It has zero conditional mean, and its time correlation defines the diffusion tensor through
\begin{equation}
\left\langle
\boldsymbol\xi_i(\rho,\phi,t)\,\boldsymbol\xi_i^{\mathrm T}(\rho,\phi,t')
\right\rangle
=
2\mathbf D(\rho,\phi)\,\delta(t-t'),
\label{eq:corr}
\end{equation}
where the average is taken over cells and time within each phase-space bin.

\begin{figure*}[t]
\centering
\includegraphics[width=\textwidth]{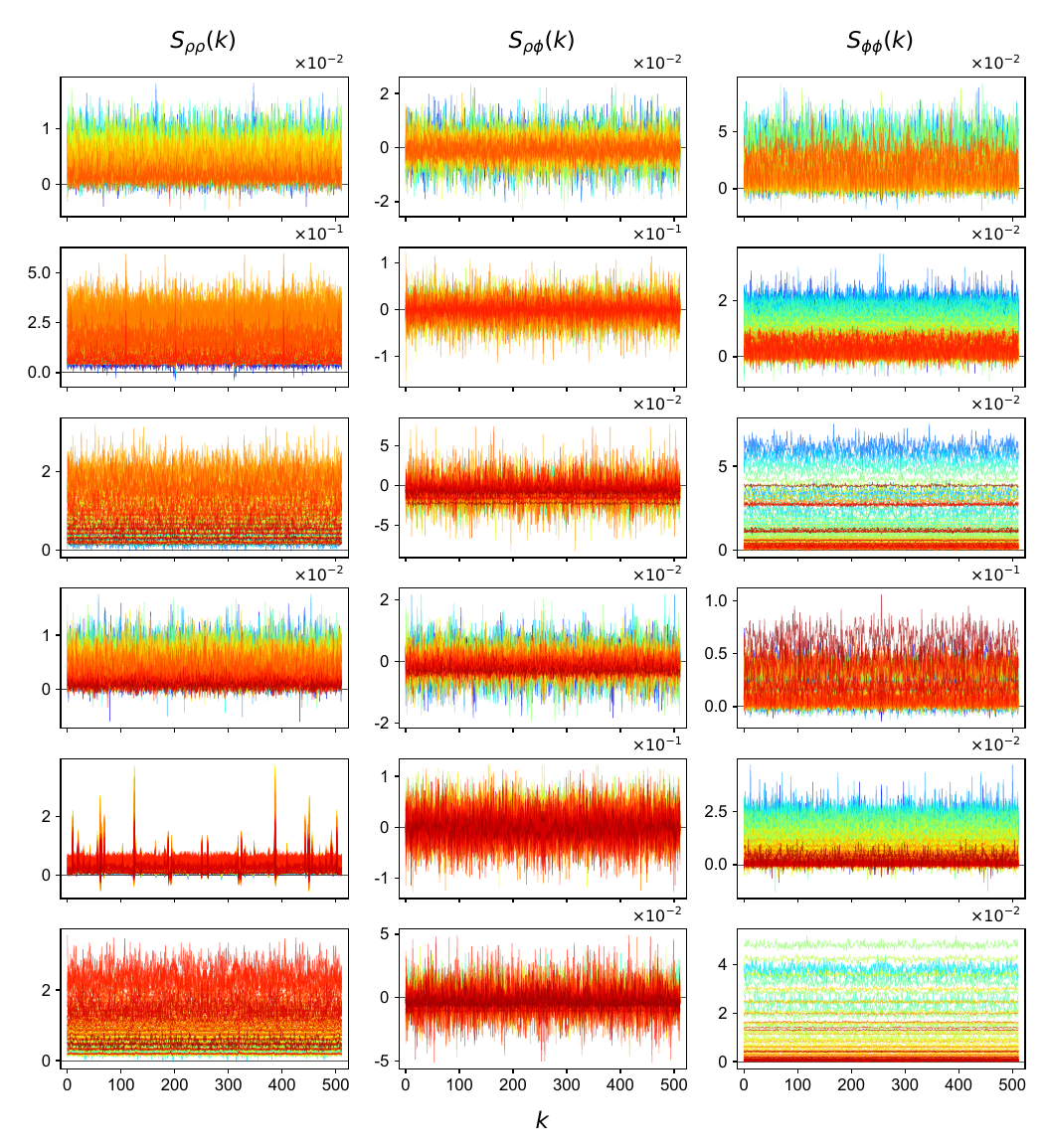}
\caption{Power spectra $S_{\rho\rho}$, $S_{\rho\phi}$, and $S_{\phi\phi}$ for intrinsic noise ($\eta_{\mathrm{in}}=0.40$, top three rows) at $\rho_0=0.5$, $1.75$, and $5$, and extrinsic noise ($\eta_{\mathrm{ex}}=0.60$, bottom three rows) at $\rho_0=0.5$, $2$, and $5$.}
\label{fig:SI_powerspectrum}
\end{figure*}

Fourier transforming this time correlation gives the spectrum $\mathbf S(\rho,\phi,\omega)$. We observe that $\mathbf S(\omega)$ is a frequency-independent spectrum (Fig.~\ref{fig:SI_powerspectrum}), revealing that the stochastic force is white noise. The diffusion tensor is therefore related to the frequency-averaged spectrum by
\begin{equation}
\mathbf D(\rho,\phi)
=
\frac12\left\langle\mathbf S(\rho,\phi,\omega)\right\rangle_\omega.
\label{eq:D}
\end{equation}

With the drift and diffusion coefficients defined above, the Fokker--Planck equation yields the corresponding stochastic Langevin equation,
\begin{equation}
\begin{aligned}
\dot\rho &= F_\rho(\rho,\phi) + \xi_\rho(t),\\
\dot\phi &= F_\phi(\rho,\phi) + \xi_\phi(t).
\end{aligned}
\label{eq:langevin}
\end{equation}

When the retained variables evolve on longer time scales than the eliminated degrees of freedom, their coarse-grained dynamics can be described by an effective Markovian Langevin equation~\cite{Esposito2012,Santillan2011,Santillan2013,Wu2021}. The drift $\mathbf F$, diffusion tensor $\mathbf D$, and steady-state distribution $P_{\mathrm{ss}}$ then provide the inputs for the driving-force decomposition, mean flux, and entropy production rate in the next section.

The cell area varies with $\rho_0$ to keep the mean particle count per cell fixed. The phase-space spacings $\delta\rho$ and $\delta\phi$ are chosen fine enough to resolve the two basins yet coarse enough that each bin retains sufficient statistics.

\section{III.\ Landscape-flux theory}
\label{sec:lft}

The nonequilibrium landscape-flux theory~\cite{Wang2008,Wang2010,Wang2015,Fang2019} decomposes the dynamics into a gradient (landscape) part and a curl (rotational flux) part. Starting from the Langevin dynamics in Eq.~\eqref{eq:langevin} with the noise statistics in Eq.~\eqref{eq:corr}, the joint probability distribution evolves according to the Fokker--Planck equation given in Sec.~II. The associated probability current is
\begin{equation}
\mathbf J=\mathbf F P-\nabla\cdot(\mathbf D P).
\label{eq:probability_current}
\end{equation}

At steady state $\nabla \cdot \mathbf{J}_{\mathrm{ss}} = 0$; this does not require $\mathbf{J}_{\mathrm{ss}} = 0$, which may instead persist as rotational currents.

The nonequilibrium potential landscape is defined as
\begin{equation}
U(\rho, \phi) = -\ln P_{\mathrm{ss}}(\rho, \phi),
\label{eq:U}
\end{equation}
representing the effective free-energy landscape of the phase space. Low-potential regions (basins and valleys) correspond to high-probability states and high-potential regions (hills and barriers) to low-probability states, so that $U$ quantifies the relative stability of the coexisting local states. By substituting the time-independent steady-state probability into Eq.~\eqref{eq:probability_current}, we obtain the steady-state probability flux
\begin{equation}
\mathbf{J}_{\mathrm{ss}}(\rho, \phi) = \mathbf{F}(\rho, \phi) P_{\mathrm{ss}} - \nabla \cdot [\mathbf{D}(\rho, \phi) P_{\mathrm{ss}}].
\label{eq:Jss}
\end{equation}
A nonzero $\mathbf{J}_{\mathrm{ss}}$ therefore signals detailed-balance breaking and drives the system away from equilibrium~\cite{XuWang2020}. To quantify its overall strength we define the mean flux magnitude
\begin{equation}
\bar J = \left\langle |\mathbf{J}_{\mathrm{ss}}| \right\rangle,
\label{eq:JP}
\end{equation}
which measures the average circulation of the steady-state probability flow over the local $(\rho,\phi)$ phase space. Together with the entropy production rate and the cross-correlation asymmetry $\Delta CC$ defined below, $\bar J$ provides the dynamical, thermodynamic, and irreversible characterization used in the main text.

Rearranging Eq.~\eqref{eq:Jss}, the deterministic driving force separates into three components with distinct physical meanings:
\begin{equation}
\mathbf{F} = \mathbf{F}_{\mathrm{grad}} + \mathbf{F}_{\mathrm{flux}} + \mathbf{F}_D = -\mathbf{D}\nabla U + \frac{\mathbf{J}_{\mathrm{ss}}}{P_{\mathrm{ss}}} + \nabla \cdot \mathbf{D}.
\label{eq:Fdecomp}
\end{equation}
The gradient force $\mathbf{F}_{\mathrm{grad}} = -\mathbf{D}\nabla U$ drives the system toward the potential basins, the flux force $\mathbf{F}_{\mathrm{flux}} = \mathbf{J}_{\mathrm{ss}}/P_{\mathrm{ss}}$ drives its rotational motion in phase space, and $\mathbf{F}_D = \nabla \cdot \mathbf{D}$ arises from the state dependence of the diffusion. In equilibrium $\mathbf{J}_{\mathrm{ss}} = 0$ and the dynamics is purely gradient-driven; the flux component $\mathbf{F}_{\mathrm{flux}}$ is what breaks detailed balance away from equilibrium.

The thermodynamic cost of the nonequilibrium dynamics is characterized by the entropy production rate ($EPR$)~\cite{Seifert2012, FangWang2020}. The Gibbs entropy of the probability distribution is
\begin{equation}
S = -\int P(\rho,\phi,t)\ln P(\rho,\phi,t)\,d\rho\,d\phi.
\label{eq:entropy}
\end{equation}
Using the Fokker-Planck equation, the rate of change of the system entropy decomposes as $\dot S = EPR - \dot S_e$, with the total entropy production rate
\begin{equation}
EPR = \int \mathbf{J}^{\mathrm{T}}\mathbf{D}^{-1}\mathbf{J}\, P^{-1}\,d\rho\,d\phi.
\label{eq:EPR_derived}
\end{equation}
Writing $\mathbf{v}(\rho,\phi,t) = \mathbf{J}(\rho,\phi,t)/P(\rho,\phi,t)$ for the local velocity of the probability flow, this becomes $EPR = \langle \mathbf{v}^{\mathrm{T}}\mathbf{D}^{-1}\mathbf{v} \rangle$, where $\langle \cdot \rangle$ is the phase-space average weighted by $P$. The conjugate entropy flow rate between the system and the environment reads
\begin{equation}
\dot S_e = \int \mathbf{J}^{\mathrm{T}}\mathbf{D}^{-1}\left(\mathbf{F} - \nabla \cdot \mathbf{D}\right)\,d\rho\,d\phi.
\label{eq:Sflow}
\end{equation}
Away from the steady state, $\dot S_e$ differs from $EPR$ by the reversible gradient contribution $\int \mathbf{J}\cdot\nabla U\,d\rho\,d\phi = \dot S$, which vanishes at steady state. In the steady state $\dot S = 0$, so that $EPR = \dot S_e \ge 0$: the system produces entropy at a rate equal to that of the entropy flowing into the environment. The steady-state $EPR$ vanishes only when $\mathbf{J}_{\mathrm{ss}}=0$, i.e. when detailed balance holds; away from equilibrium it quantifies the thermodynamic cost of the non-conservative curl flux and provides a global measure of the time-reversal asymmetry of the coarse-grained dynamics.

\section{IV.\ Landscape evolution across density}
\label{sec:landscape}

\begin{figure*}[t]
\centering
\includegraphics[width=\textwidth]{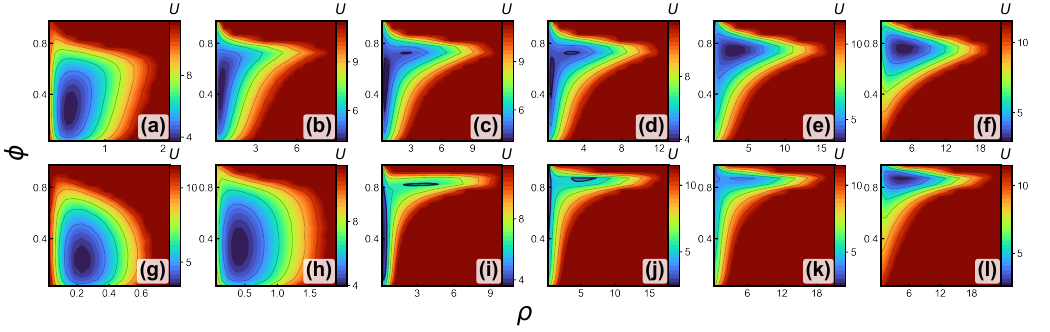}
\caption{Nonequilibrium potential landscapes $U(\rho,\phi)$ at increasing total density $\rho_0$ under intrinsic noise (top row, panels a--f; $\eta_{\mathrm{in}}=0.4$) and extrinsic noise (bottom row, panels g--l; $\eta_{\mathrm{ex}}=0.6$). The total densities are $\rho_0=0.5$, $1.0$, $1.5$, $1.75$, $3.0$, and $5.0$ for the top row and $\rho_0=0.25$, $0.5$, $0.75$, $2.0$, $3.0$, and $5.0$ for the bottom row. The colorbar indicates the landscape value $U$.}
\label{fig:SI_landscape}
\end{figure*}

The main text reports the potential landscapes at three representative densities for each noise type. To substantiate the continuous-like versus abrupt character of the basin transitions, we provide here the landscape sequence at a denser set of densities around the two basin boundaries, for both intrinsic and extrinsic noise (Fig.~\ref{fig:SI_landscape}).

As $\rho_0$ increases, the landscape evolves from a single disordered basin located in the low-$\phi$ region, to two coexisting basins (the dilute disordered gas and the dense ordered band)~\cite{Kursten2020, Mangeat2020, Clusella2021}, and finally to a single ordered basin located in the high-$\phi$ region. Under intrinsic noise, the appearance and disappearance of the two-basin landscape develop smoothly over a broad density interval, consistent with a continuous-like transition. Under extrinsic noise, both basin transitions are confined to a narrow density range and are therefore markedly sharper. This extended density resolution visualizes the contrast directly and complements the $\phi_0$-$\rho_0$ curves in the main text.

\section{V.\ Cross-correlation asymmetry $\Delta CC$}
\label{sec:cc}

The cross-correlation asymmetry $\Delta CC$ used in the main text quantifies time-reversal symmetry breaking from the two-point cross-correlation functions of the density and order-degree time series:
\begin{align}
C_{\rho\phi}(\tau)&=\langle \rho_m(t)\,\phi_m(t+\tau) \rangle, \nonumber\\
C_{\phi\rho}(\tau)&=\langle \rho_m(t+\tau)\,\phi_m(t) \rangle,
\label{eq:C_rhophi}
\end{align}
where $m$ indexes the position-space cells, $t$ is the sampling time, and the average $\langle\cdot\rangle$ is taken over all cells, all time origins, and the ensemble. The asymmetry measure reads
\begin{equation}
\Delta CC = \sqrt{\left\langle \left[ C_{\rho\phi}(\tau) - C_{\phi\rho}(\tau) \right]^{2} \right\rangle_{\tau}},
\label{eq:deltaCC}
\end{equation}
where $\langle\cdot\rangle_\tau$ denotes the average over lag steps $\tau$ up to a maximum lag $\tau_{\max}$ that covers the relaxation of the coarse-grained variables.
In a time-reversal-symmetric system the cross-correlation is invariant under exchanging the two observables together with the sign of $\tau$, so that $C_{\rho\phi}(\tau) = C_{\phi\rho}(\tau)$ for all $\tau$ and $\Delta CC=0$. A nonzero $\Delta CC$ therefore indicates directional coupling between local density and local order and exposes time-reversal symmetry breaking~\cite{Borthne2020}.

\begin{figure*}[t]
\centering
\includegraphics[width=\textwidth]{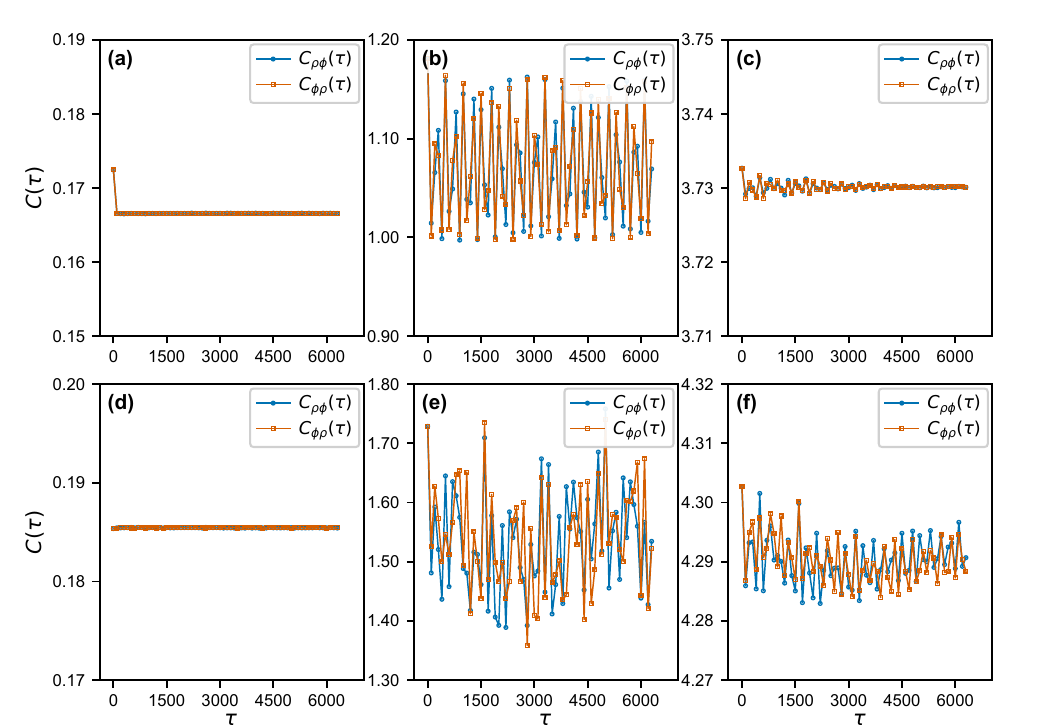}
\caption{Raw cross-correlation functions $C_{\rho\phi}(\tau)$ (blue) and $C_{\phi\rho}(\tau)$ (orange) versus the lag $\tau$ at representative densities in the disordered, band, and ordered phases under intrinsic noise (top row, $\eta_{\mathrm{in}}=0.4$; $\rho_0=0.5$, $1.75$, and $5.0$) and extrinsic noise (bottom row, $\eta_{\mathrm{ex}}=0.6$; $\rho_0=0.5$, $2.0$, and $5.0$).}
\label{fig:SI_cc}
\end{figure*}

Figure~\ref{fig:SI_cc} presents the raw cross-correlation functions $C_{\rho\phi}(\tau)$ and $C_{\phi\rho}(\tau)$ at representative densities in the disordered, band, and ordered phases for both noise types. The forward and backward cross-correlations nearly coincide in the disordered and ordered phases and deviate most strongly in the band phase region, giving rise to the $\Delta CC$ peak reported in the main text (Fig.~3 therein).

\section{VI.\ Optimal transition paths}
\label{sec:paths}

The optimal transition paths between phases are obtained from the path-integral formulation of the Langevin dynamics~\cite{Wang2010}. Each path is weighted by an effective action, and the optimal path minimizes the action with the endpoints fixed at the minima of the landscape $U(\rho,\phi)$ that locate the disordered and ordered basins. The forward path is the one along which the system switches from the disordered basin to the ordered basin, and the backward path the reverse.

Because $\mathbf{D}$ enters the Fokker--Planck equation~\eqref{eq:FP} inside the second derivative, the Langevin equation~\eqref{eq:langevin} is interpreted in the Ito sense, with $\mathbf{F}$ the conditional mean drift and $\mathbf{D}$ the covariance of the residual noise [Eqs.~\eqref{eq:F}--\eqref{eq:corr}]. For a path $\mathbf{x}(t)$ from $\mathbf{x}_i$ at $t=0$ to $\mathbf{x}_f$ at $t=t_f$, the path probability is
\begin{equation}
P(\mathbf{x}_f,t_f|\mathbf{x}_i,0)=\int\mathcal{D}[\mathbf{x}]\,e^{-S[\mathbf{x}]},
\label{eq:path_probability}
\end{equation}
with the effective Onsager--Machlup action
\begin{equation}
S[\mathbf{x}]=\int_0^{t_f}dt\left[\frac{1}{4}\left(\dot{\mathbf{x}}-\mathbf{F}\right)^{\mathrm{T}}\mathbf{D}^{-1}(\mathbf{x})\left(\dot{\mathbf{x}}-\mathbf{F}\right)+\frac{1}{2}\,\mathrm{div}_{\mathbf{D}}\mathbf{F}(\mathbf{x})\right],
\label{eq:action}
\end{equation}
where
\begin{equation}
\mathrm{div}_{\mathbf{D}}\mathbf{F}\equiv\nabla\cdot\mathbf{F}-\mathbf{F}^{\mathrm{T}}\mathbf{D}^{-1}\left(\nabla\cdot\mathbf{D}\right),\qquad\left(\nabla\cdot\mathbf{D}\right)_i=\sum_j\partial_j D_{ij}.
\label{eq:divD}
\end{equation}
The first term is the Gaussian weight of the noise realization. The second term contains the Jacobian of the noise-to-path transformation together with the noise-induced contribution from the state dependence of $\mathbf{D}$; it reduces to $\frac{1}{2}\nabla\cdot\mathbf{F}$ when $\mathbf{D}$ is constant. The determinant prefactor is absorbed in the path-integral measure.

Because the transition time is not fixed by the steady-state data, we minimize $S$ in the Hamilton--Jacobi (length-space) representation. Defining
\begin{equation}
V_{\mathrm{eff}}(\mathbf{x})=\frac{1}{4}\mathbf{F}^{\mathrm{T}}\mathbf{D}^{-1}\mathbf{F}+\frac{1}{2}\,\mathrm{div}_{\mathbf{D}}\mathbf{F},\qquad E_{\mathrm{eff}}=-\min\left[V_{\mathrm{eff}}(\mathbf{x}_A),V_{\mathrm{eff}}(\mathbf{x}_B)\right],
\label{eq:Veff}
\end{equation}
and discretizing the path into $N$ segments $\{\mathbf{x}_n\}_{n=0}^{N}$ with $\mathbf{x}_0=\mathbf{x}_A$ and $\mathbf{x}_N=\mathbf{x}_B$, the action reads
\begin{equation}
S_{\mathrm{HJ}}=\sum_{n=0}^{N-1}\left[\sqrt{E_{\mathrm{eff}}+V_{\mathrm{eff}}(\mathbf{x}_{n+1/2})}\,\Delta l'_n-\frac{1}{2}F'_n\Delta l_n\right]+\lambda\mathcal{P},
\label{eq:SHJ}
\end{equation}
with $\mathbf{x}_{n+1/2}=(\mathbf{x}_n+\mathbf{x}_{n+1})/2$,
\begin{align}
\Delta l_n&=\left|\mathbf{x}_{n+1}-\mathbf{x}_n\right|,\\
\Delta l'_n&=\sqrt{\left(\mathbf{x}_{n+1}-\mathbf{x}_n\right)^{\mathrm{T}}\mathbf{D}^{-1}(\mathbf{x}_{n+1/2})\left(\mathbf{x}_{n+1}-\mathbf{x}_n\right)},\\
F'_n&=\frac{\left(\mathbf{x}_{n+1}-\mathbf{x}_n\right)^{\mathrm{T}}\mathbf{D}^{-1}(\mathbf{x}_{n+1/2})\,\mathbf{F}(\mathbf{x}_{n+1/2})}{\Delta l_n},
\end{align}
and
\begin{equation}
\mathcal{P}=\sum_{n=0}^{N-1}\left(\Delta l_n-\langle\Delta l\rangle\right)^2,\qquad\langle\Delta l\rangle=\frac{1}{N}\sum_{n=0}^{N-1}\Delta l_n.
\label{eq:penalty}
\end{equation}
The endpoints $\mathbf{x}_A$ and $\mathbf{x}_B$ are the two local minima of $U(\rho,\phi)$ that correspond to the disordered and ordered phases at the chosen $\rho_0$; the intermediate points and the total time are free. The path is initialized as an interpolated path between the two minima and optimized by a greedy random-scan Monte Carlo search. At each trial a random sub-interval of the path is selected, each interior point in that interval is displaced along one randomly chosen coordinate by at most half a grid spacing, and the move is accepted only if the penalized action decreases. The penalty enforces nearly uniform segment lengths. Convergence is monitored through the residual action gradient at the interior points and the condition number of $\mathbf{D}$ along the path, and the optimized path is compared with the initial path and with a monotonic straight path. The drift and diffusion are evaluated from the tabulated grid fields at the midpoint of each segment.

Equation~\eqref{eq:Fdecomp} separates the drift into a gradient part and a flux part. At the level of the stochastic process, $\mathbf{J}_{\mathrm{ss}}=0$ implies detailed balance: a path and its time reverse then carry the same statistical weight up to endpoint factors, and the two optimal paths coincide. A nonzero $\mathbf{J}_{\mathrm{ss}}$ breaks this equivalence and generally separates them. For a gradient drift with constant $\mathbf{D}$, the second term of Eq.~\eqref{eq:SHJ} telescopes to the endpoint difference $\frac{1}{2}[U(\mathbf{x}_B)-U(\mathbf{x}_A)]$ and drops out of the Euler--Lagrange equation, which is the explicit form of this statement; when the non-gradient part $\mathbf{F}_{\mathrm{flux}}$ is present, the same term becomes path dependent and acts as a curl force in that equation. Because $\mathbf{D}$ is state dependent here, the discretized functional also retains a metric contribution $-\frac{1}{2}\int(\mathbf{D}^{-1}\nabla\cdot\mathbf{D})\cdot d\mathbf{x}$ that is not a pure endpoint difference. The computed separation is therefore a signature of the flux-induced irreversibility of the inferred dynamics rather than an exact scalar measure of $|\mathbf{J}_{\mathrm{ss}}|$: the gradient part and the anisotropy of $\mathbf{D}$ also shape the paths.

\begin{figure*}[t]
\centering
\includegraphics[width=\textwidth]{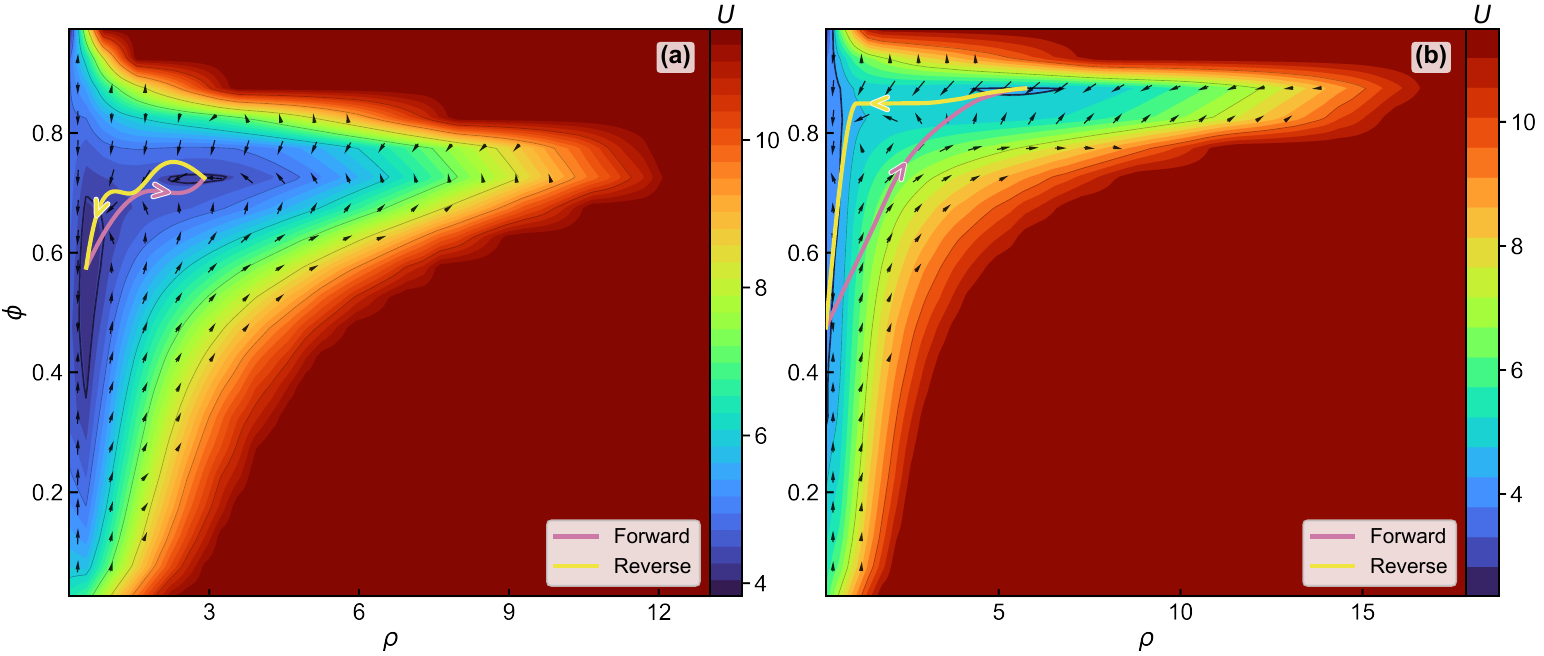}
\caption{Optimal transition paths between ordered and disordered phases in the local $(\rho,\phi)$ phase space for intrinsic noise (a) and extrinsic noise (b), at the total densities $\rho_0=1.75$ and $\rho_0=2.0$, respectively. The forward path (pink line) represents the most probable trajectory from the disordered phase to the ordered phase, while the backward path (yellow line) represents the opposite transition.}
\label{fig:SI_paths}
\end{figure*}

Figure~\ref{fig:SI_paths} shows the optimal transition paths for intrinsic noise ($\rho_0=1.75$) and extrinsic noise ($\rho_0=2.0$). In both cases the forward and backward paths remain distinct. The separation is more pronounced under extrinsic noise, consistent with the stronger time-reversal asymmetry reported by $\Delta CC$.

\section{VII.\ Phase diagram classification}
\label{sec:phasediag}

In the main text, the phase diagrams in the $(\rho_0,\eta)$ parameter space are obtained by classifying each sampled steady-state particle configuration into the disordered, band, or ordered phase~\cite{Chate2020}. The classification is based on the structure of the steady-state particle configurations, cross-checked against the basin structure of the potential landscape. In the disordered phase the particles form no coherent spatial structure and the total order degree $\phi_0$ is low. In the band phase a dense ordered band propagates through and coexists with a dilute disordered gas. In the ordered phase the system reaches a high total order degree, with most particles moving in a common direction. The lower and upper boundaries $\rho_{0,\mathrm{lower}}$ and $\rho_{0,\mathrm{upper}}$ of the band phase region are read off from the configurations, and the band width is defined as $\Delta\rho_0 = \rho_{0,\mathrm{upper}}-\rho_{0,\mathrm{lower}}$.

\begin{figure*}[t]
\centering
\includegraphics[width=\textwidth]{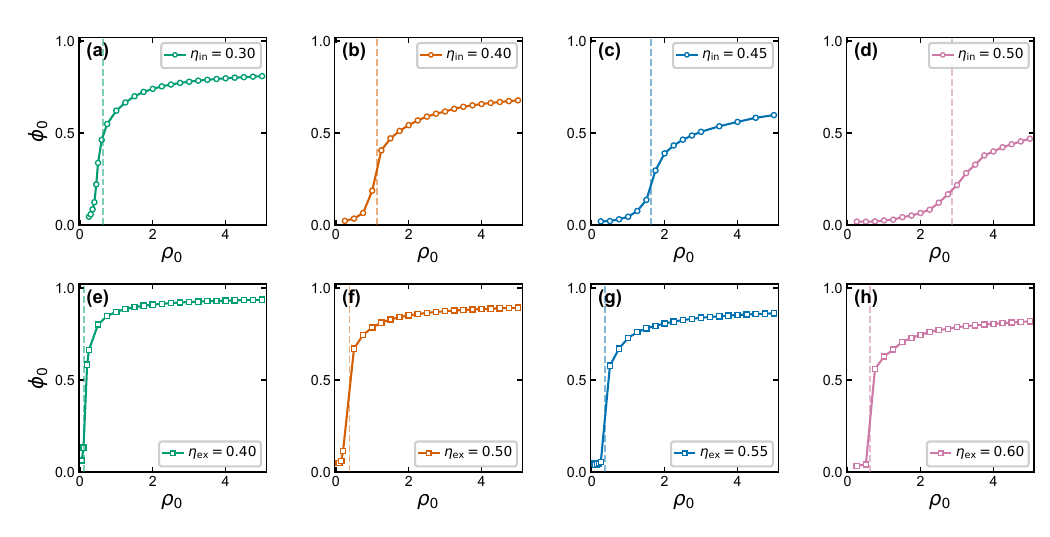}
\caption{Total order degree $\phi_0$ versus total density $\rho_0$ at several noise intensities for intrinsic noise (a--d, $\eta_{\mathrm{in}}=0.30$, $0.40$, $0.45$, and $0.50$) and extrinsic noise (e--h, $\eta_{\mathrm{ex}}=0.40$, $0.50$, $0.55$, and $0.60$). Dashed vertical lines mark the transition densities.}
\label{fig:SI_phirho}
\end{figure*}

Figure~\ref{fig:SI_phirho} shows the $\phi_0$-$\rho_0$ curves at several noise intensities, which correspond to fixed-$\eta$ cuts through the phase diagrams in Fig.~4(a,b) of the main text. Under intrinsic noise [Fig.~\ref{fig:SI_phirho}(a--d)], $\phi_0$ grows continuously with $\rho_0$ over a broad density interval. Under extrinsic noise [Fig.~\ref{fig:SI_phirho}(e--h)], $\phi_0$ stays low up to a well-defined density and then jumps abruptly, so that the transition is discontinuous.

\clearpage

\begin{figure*}[t]
\centering
\includegraphics[width=\textwidth]{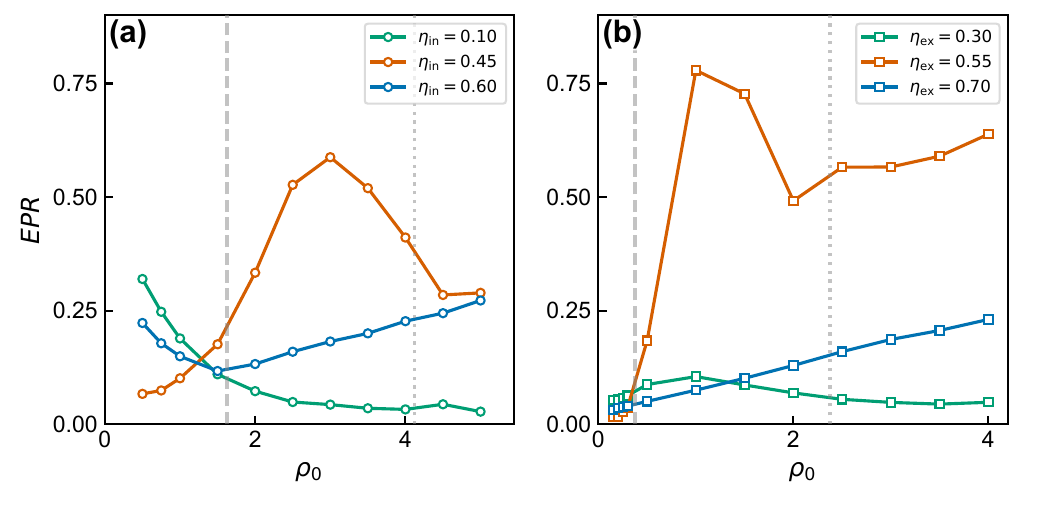}
\caption{Entropy production rate $EPR$ versus total density $\rho_0$ under intrinsic noise (a) and extrinsic noise (b), at the noise intensities shown in the legends.}
\label{fig:SI_EPR}
\end{figure*}

\FloatBarrier

\clearpage
\bibliography{ref.bib}